\documentstyle[12pt]{article}

\newcommand{\sect}[1]{\setcounter{equation}{0}\section{#1}}
\newcommand{\subsect}[1]{\subsection{#1}}

\renewcommand{\theequation}{\arabic{section}.\arabic{equation}}

\def\be{\begin{equation}}
\def\ee{\end{equation}}
\def\bea{\begin{eqnarray}}
\def\eea{\end{eqnarray}}

\def\p{\varepsilon}

\def\k{\kappa}
\def\Ck{{\rm{\ \!C\/}}}
\def\Sk{{\rm{\ \!S\/}}}

\def\co{\Delta}

\def\>#1{{\bf #1}}
\def\1{\'{\i}}
\def\R{\rm I\kern-.2em R}
\def\back{\!\!\!\!\!\!}

\def\JJ{\widetilde{J}}

\def\diag{{\mbox{diag\,}}}

\def\tfrac#1#2{{\scriptstyle{\frac{#1}{#2}}}}

\begin{document}

\thispagestyle{empty}

\hfill\today

\vspace{2cm}

\begin{center}
{\LARGE{\bf{Lie bialgebra contractions  }}}

{\LARGE{\bf{   and quantum deformations  of
 }}}

{\LARGE{\bf{ quasi-orthogonal algebras}}}
\end{center}

\bigskip\bigskip\bigskip

\begin{center}
A. Ballesteros$^{1}$, N. A. Gromov$^{2}$, F.J. Herranz, M.A. del
Olmo and M. Santander
\end{center}

\begin{center}
\em

{ { Departamento de F\1sica Te\'orica, Universidad de Valladolid } \\
E-47011, Valladolid, Spain }

{ { {}$^{1}$ Departamento de F\1sica, Universidad de Burgos}
\\  Facultad de CYTA y Ciencias Qu\1micas,  E-09003, Burgos, Spain}

{ $^{2}$ In absence from
   Department of Mathematics,  Komi Scientific Centre \\
   167000 Syktyvkar, Russia}
   \end{center}
\rm

\bigskip

\bigskip\bigskip\bigskip

\begin{abstract}
Lie bialgebra contractions are introduced and classified.
A non-degenerate coboundary bialgebra structure is implemented into
all pseudo-orthogonal algebras $so(p,q)$
starting from the one corresponding to
$so(N+1)$. It allows to introduce a   set of
Lie bialgebra contractions which leads to  Lie bialgebras
of quasi-orthogonal algebras. This construction is explicitly given for
the cases $N=2,3,4$. All
Lie bialgebra contractions studied in this paper define Hopf algebra
contractions for the Drinfel'd-Jimbo deformations $U_z so(p,q)$. They are
explicitly used to generate new non-semisimple quantum algebras as it is
the case for the Euclidean, Poincar\'e and Galilean algebras.

  \end{abstract}

\newpage

\sect{Introduction}

Several non-semisimple Lie groups play an important role in Physics, for
instance, the Poincar\'e and Galilei ones; they can be got starting from
semisimple groups by means of a sequence of contractions \cite{IW}.
The current
interest in quantum deformations of Lie algebras raised the
extension of the idea
of contraction from Lie algebras to their quantum analogues
taking into account the bearing of the contraction
on the deformation parameter \cite{CGST}. In this way,
the number of different
possible
contractions to perform on a given algebra increases
significantly, as well as
the difficulties encountered  to analyse the convergency
properties of each of
them.

On the other hand, the main underlying structure of a quantum algebra
is the Lie bialgebra that gives the first order term in the
deformation \cite{Dri,Pres}. Higher order terms can be,
in principle, obtained
by a consistency method, and the classification problem for
quantum deformations is rather simplified by taking into account
this fact \cite{ZakrLor}. The aim of this paper is to show that the
information related to a given Hopf algebra contraction can be
extracted with
much less effort from the underlying Lie bialgebra, thus providing a
simplified approach to a (constructive) classification of Hopf algebra
contractions.

This program is developed here for  the non-degenerate
(or standard) coboundary
Lie bialgebras linked to the quantum  orthogonal algebras $so(N+1)$;
they are generated by an $r$--matrix which satisfies the
modified classical
Yang--Baxter equation (YBE). Firstly, we endow the pseudo-orthogonal
algebras
$so(p,q)$ with this  bialgebra structure and afterwards we study a
set of
Lie bialgebra contractions which provides quasi-orthogonal bialgebras.

  The main concepts   about (coboundary) Lie bialgebras and their
  contractions
are established in section 2. The cases $N=2 ,3, 4$ are fully
analysed in sections 3, 4 and 5, respectively.  We explicitly study
all the
possible choices of the behaviour of the deformation parameter
under the chosen
contractions, and classify the divergencies they produce both
in the classical
$r$--matrix and in the bialgebra mapping $\delta$. We study separately
these objects,
since starting from a coboundary Lie bialgebra a contraction
can either produce
a  coboundary bialgebra (both $r$ and $\delta$ do not diverge)
or  a (right)
bialgebra that   is not a coboundary ($r$ diverges but $\delta$ is well
defined). Therefore, a separate analysis of the
behaviour of $r$ and $\delta$
under contraction makes more clear the link between
non-semisimple algebras and
non-coboundary structures.

\sect{Lie bialgebras and their contractions}

Firstly, we present the basic concepts that we shall need in order to
introduce Lie bialgebra contractions. Basic facts
about quantum algebras can
be found in references \cite{Dri,Pres,Ji,FRT,Maj,Tjin,Dobrev}.

\subsect{Lie bialgebras}

\noindent {\bf Definition 2.1.}
\begin{em}
A Lie bialgebra $(g,\eta)$ is a Lie algebra $g$ endowed with a
cocommutator $\eta:g\to g\otimes g$ such that

\noindent i) $\eta$ is a 1--cocycle, i.e.,
\be
\eta([x,y])=[\eta(x),\, 1\otimes y+ y\otimes 1] +
[1\otimes x+ x\otimes 1,\, \eta(y)], \quad \forall x,y\in
g. \label{bc}
\ee

\noindent ii) The dual map $\eta^\ast:g^\ast\otimes g^\ast \to
g^\ast$ is a Lie bracket on $g^\ast$.
\end{em}

\noindent {\bf Definition 2.2.}
\begin{em}
A Lie bialgebra $(g,\eta)$ is called a coboundary bialgebra if
there exists an element $\rho\in g\otimes g$ called $r$--matrix, such that
\end{em}
\be
\eta(x)=[1\otimes x + x \otimes 1,\,  \rho], \qquad  \forall x\in
g.
\label{bd}
\ee

It can be easily shown that the map (\ref{bd}) defined by means
of an arbitrary $\rho$ is a Lie bialgebra if and only if the
symmetric part of $\rho$ is a $g$--invariant element of $g\otimes
g$ and the Schouten bracket
\be
[\rho,\rho]:=[\rho_{12},\rho_{13}] + [\rho_{12},\rho_{23}] +
[\rho_{13},\rho_{23}]
\label{bdc}
\ee
is a $g$--invariant element of $g\otimes g \otimes g$ ($\rho$
fulfills the modified classical YBE: $ad_g
[\rho,\rho]=0$). Here $\rho_{12}=\rho\otimes 1$, and the same convention is
taken for $\rho_{13}$ and $\rho_{23}$.
In fact, we shall consider skew-symmetric
$r$--matrices, and we shall denote coboundary Lie bialgebras in the form

$(g,\eta(\rho))$.

\noindent {\bf Definition 2.3.}
\begin{em}
Two Lie bialgebras $(g,\eta_1)$ and $(g,\eta_2)$ are said
to be equivalent if there exists a Lie algebra automorphism
$D$ of $g$ such that $\eta_2=(D^{-1}\otimes D^{-1}) \circ \eta_1
\circ D$.
\end{em}

As a consequence, two $r$--matrices
$\rho_1$ and $\rho_2$ will be considered
as equivalent if the bialgebras
$(g,\eta_1(\rho_1))$ and $(g,\eta_2(\rho_2))$
generated by them are equivalent according to Def.\ 2.3.

Let us recall that a quantum universal enveloping algebra (QUEA) $A=U_z
g$ is a quantization of a Lie bialgebra $(g,\eta)$ in the following
sense: if we write the coproduct $\Delta:A\to A\otimes A$  as a
formal power series in the deformation parameter $z$
\be
\Delta=\sum_{k=0}^{\infty} \Delta_{(k)}=\sum_{k=0}^{\infty}
z^k \eta_{(k)}, \label{be}
\ee
the classical limit condition $A/zA\equiv U g$ implies that the
mapping defined by
\be
\eta :=(\Delta_{(1)} - \sigma\circ
\Delta_{(1)})\quad \mbox{mod}\,z ,
\label{bf}
\ee
is a Lie bialgebra mapping (here, $\sigma(a\otimes
b):=b\otimes a$). If $\Delta_{(1)}$ is skew-symmetric,
we have that $\eta =2 \eta_{(1)}$. Moreover, higher order
terms in the coproduct (\ref{be}) can be reconstructed in terms
of the first order one by solving the coassociativity condition $(1\otimes
\Delta)\Delta=(\Delta\otimes 1)\Delta$ order by order in $z$, which reads:
\be
\sum_{n=0}^{k}   (\Delta_{(n)}\otimes 1 -
1 \otimes \Delta_{(n)})\Delta_{(k-n)}=0.
\label{bh}
\ee

\subsect{Contraction of Lie bialgebras}

We first recall the definition of Lie
algebra contraction in a general setting \cite{Sal,Doo}. Hereafter, $g$
will be a finite dimensional  Lie algebra.

\noindent {\bf Definition 2.4.}
\begin{em}
A Lie algebra $g'$ is a contraction of another
Lie algebra $g$ (with the same underlying vector space $V$) if
there exists a oneparametric family of  Lie algebra automorphisms

$\phi_\p:g\to g$ such that the limit $\lim_{\p\to 0}
{\phi_\p^{-1}[\phi_\p(X),\phi_\p(Y)]}$ of the Lie bracket in $g$
gives the Lie bracket $[X,Y]'$ in $g'$. \end{em}

The mappings $\phi_\p$ can be interpreted, as long as $\p\ne 0$, as an
embedding of the Lie generators of $g'$ within $g$.
Relations among the former ones are computed as embedded in
$g$, and turned back to $g'$ by using $\phi_\p^{-1}$
and then making the limit $\p \to 0$. We will assume that the automorphisms

$\phi_\p$ have a polynomial dependence   on $\p$, as it is the case in most
physical examples of contractions. The contraction of the Lie
bracket $[\, , \,]:g\otimes g\rightarrow g$ so defined can be
generalized to the bialgebra mapping
$\eta:g\rightarrow g\otimes g$ and to the $r$--matrix $\rho\in g\otimes g$
as follows:

\noindent {\bf Proposition 2.5.}
\begin{em}
Let $(g,\eta)$ be a Lie bialgebra and let $g'$ be a
contraction of $g$ defined by
the   mappings  $\phi_\p$. If $n$ is any real number such that the
limit
\be
\eta':=\lim_{\p \to 0}{\p^n
(\phi_\p^{-1}\otimes\phi_\p^{-1})\circ\eta\circ
\phi_\p},
\label{bi}
\ee
exists, then $(g',\eta')$ is a Lie bialgebra. Furthermore, there exists a
single minimal fixed value $f_0$ of $n$ such that for $n\ge f_0$ the limit
(\ref{bi}) exists and for $n>f_0$ the limit is zero.
\end{em}

For a given family of automorphisms $\phi_\p$ defining
the contraction of the
Lie algebra $g\to g'$, we get from Prop.\ 2.5 a family of Lie bialgebras
$(g',\eta')$ parametrized by the real number $n\ge f_0$,
which can be naturally
considered as contractions of $(g,\eta)$. Note however that $\eta'$ is a
non-trivial cocommutator only when $n=f_0$. Therefore, the following
definitions make sense:

\noindent {\bf Definition 2.6.}
\begin{em}
The Lie bialgebra $(g',\eta')$ is said
to be a  Lie bialgebra contraction  (LBC) of $(g,\eta)$ if there
exists a contraction from $g$ to $g'$ described by a family
$\phi_\p$ of Lie
algebra automorphisms and a number $n$ such that $\eta'$ is given by the
limit  (\ref{bi}). We denote such a contraction by the pair $(\phi_\p,n)$.
\end{em}

\noindent {\bf Definition 2.7.}
\begin{em}
The minimal value   $f_0$ of $n$
will be called the  fundamental
contraction constant  of the Lie bialgebra $(g,\eta)$ associated to the
family $\phi_\p$.
\end{em}

\noindent {\bf Definition 2.8.}
\begin{em}
The LBC with the minimal value $f_0$ of $n$ $(\phi_\p,f_0)$ is said to be
the  fundamental LBC  of $(g,\eta)$ associated to the mappings
$\phi_\p$. \end{em}

The preceding discussion applies to any Lie bialgebra, not necessarily a
coboundary one. But for a coboundary Lie bialgebra it is
rather natural to study
directly the contraction of the associated $r$--matrix. More specifically:

\noindent {\bf Proposition 2.9.}
\begin{em}
Let $(g,\eta(\rho))$ be a coboundary Lie bialgebra and let $g'$ be a
contraction of $g$ defined by the   mappings  $\phi_\p$. If $n$ is a real
number such that the limit
\be
\rho':=\lim_{\p \to 0}{\p^n
(\phi_\p^{-1}\otimes \phi_\p^{-1}) (\rho)},
\label{bk}
\ee
exists, then $(g',\eta'(\rho'))$ is a coboundary Lie bialgebra
where $\eta'$
is obtained by applying (\ref{bd}).
Moreover, there exists a single minimal fixed value $c_0$ of $n$
such that for
$n\ge c_0$ the limit (\ref{bk}) exists and for $n>c_0$ the limit is zero.
\end{em}

Obviously,  the coboundary bialgebra $(g',\eta'(\rho'))$ is a
contraction of  $(g,\eta(\rho))$ in the sense of Def. 2.6.
On the other hand, Defs.\
2.7 and 2.8 can be suitably modified leading to:

\noindent {\bf Definition 2.10.}
\begin{em}
The minimal value $c_0$ of $n$
will be called the  coboundary
contraction constant  of the Lie bialgebra $(g,\eta(\rho))$ associated to
the family $\phi_\p$.
\end{em}

\noindent {\bf Definition 2.11.}
\begin{em}
The LBC with the minimal value $c_0$ of $n$   $(\phi_\p,c_0)$ is said to be
the  coboundary LBC  of $(g,\eta(\rho))$ associated to the
mappings $\phi_\p$. \end{em}

It is clear that $f_0\le c_0$ always hold (cfr.\ Prop.\ 2.5);
the important point
to be emphasized here is that both limits (\ref{bi}) and
(\ref{bk}), being of a
different nature, should be analysed separately and do not necessarily lead to
equal values for the fundamental and coboundary contraction
constants of a given
coboundary bialgebra. When $f_0=c_0$, the contraction
$(\phi_\p,n)$ with $n=c_0=f_0$ is called a \begin{em}
``fundamental coboundary LBC"\end{em} (both the contracted
cocommutator and
$r$--matrix are not trivial). If $f_0<c_0$ the  LBC with coboundary
contraction constant $c_0$ is {\it not} a fundamental one, because  the
contracted cocommutator is zero and $r'$ will be either
trivial or equivalent
to the trivial one; hence, in this situation the  LBC with fundamental
contraction constant $f_0$ is {\it not} a coboundary one.

In general, it is necessary to allow for $n$ in order to
ensure the convergency of the limit (\ref{bi}). However, if we
consider Lie bialgebras as generating objects for quantum
algebras, this fact can be interpreted in a
different way (as a kind of ``renormalization") by
including the deformation
parameter $z$ within the Lie cocommutator, that we shall
denote as $\delta(X):=z\, \eta(X)$. This can be done provided
we are able to find a homomorphism $D$ giving
the equivalence between $\delta$ and $\eta$ (see Def.\ 2.3.; the mere
multiplication of all generators of $g$ by $z$ gives the simplest form of
$D$). In this sense, the renormalization factor $\p^n$
can be associated to
a transformation law of the deformation parameter
\be
w=\phi_\p(z):=\p^{-n}z, \label{bka}
\ee
where $w$ would be the new (contracted) deformation
parameter. The contracted structure can be now defined without any factor
$\p^n$:
\bea
\delta'&:=&\lim_{\p \to 0}{
(\phi_\p^{-1}\otimes\phi_\p^{-1})\circ\delta\circ
\phi_\p}=\lim_{\p \to 0}{z\,
(\phi_\p^{-1}\otimes\phi_\p^{-1})\circ\eta\circ
\phi_\p}\cr
&=&\lim_{\p \to 0}{w \,\p^{n}\,
(\phi_\p^{-1}\otimes\phi_\p^{-1})\circ\eta\circ
\phi_\p}=w \,\eta' . \label{bkc}
\eea

Obviously, if  the Lie bialgebra $(g,\eta)$ is a coboundary
one generated by
$\rho$, then $(g,\delta)$ will be also a coboundary Lie
bialgebra generated by
$r:=z\rho$. The very same extension of the action of the
mappings  $\phi_\p$
given by (\ref{bka}) leads to the following modification of (\ref{bk}):
\be
r':=\lim_{\p \to 0}{
(\phi_\p^{-1}\otimes\phi_\p^{-1})(r)}
=\lim_{\p \to 0} w\, \p^n   {
(\phi_\p^{-1}\otimes\phi_\p^{-1})(\rho)}=w\, \rho'.
\label{bl}
\ee

This behaviour
of the deformation parameter  was soon
discovered as a condition to obtain non-semisimple quantum
algebras by contraction \cite{CGST}; moreover it provided a dimensional
interpretation of the quantum parameter \cite{BH4D}.
At this respect, the LBC
framework   we have just developed arises as relevant, since for all the
cases that have been worked out, the existence of a fundamental LBC
$(\phi_{\p},f_0)$ is a {\it sufficient} condition for
the convergency of the
entire Hopf structure under the Hopf algebra contraction defined by
\be
\Delta':=\lim_{\p \to 0}{
(\phi_\p^{-1}\otimes\phi_\p^{-1})\circ\Delta\circ
\phi_\p}. \label{bld}
\ee
This fact seems to be rather general: for a given Lie
algebra contraction, the
fundamental LBC defines in a unique way the transformation law of $z$ (the
change of generators is alocady given by the classical contraction
$\phi_{\p}$) and ensures the first order deformation to be well-defined.
Afterwards, the coassociativity constrain enters, and propagates
the convergency
of this first order term to higher order ones.

In the sequel, this result is shown to be very useful, since the systematic
classification of LBC's allows us to find some new quantum non-semisimple
algebras   by considering all possible transformations of
$z$ that still define
right LBC's. Non-coboundary structures will appear in a natural way in this
context, and a global overview of all possible contractions of a given Hopf
algebra is obtained.

\subsect{Quasi-orthogonal algebras as contractions from $so(p,q)$ algebras}

 The purpose of this paper is to analyse the contraction scheme sketched in the
previous section for a specific particular set of Lie bialgebras,
which include
pseudo-orthogonal and quasi-orthogonal bialgebras.
For any $N=2, 3, \dots$, let us
consider the real Lie algebra with $N(N+1)/2$ generators $J_{ab}$, $a<b;
(a,b=0,1,\dots, N)$, and commutation relations: \be
[J_{ab}, J_{ac}] = \k_{ab} J_{bc}, \quad
[J_{ab}, J_{bc}] = -J_{ac},  \quad
[J_{ac}, J_{bc}] = \k_{bc} J_{ab}, \quad a<b<c,
\label{bp}
\ee
(those commutators involving four different indices are equal to zero),
where
the structure constants $\k_{ab}$ depend on $N$ real numbers $\k_1, \dots
,\k_N$:
\be
 \k_{ab}= \prod_{s=a+1}^b\k_s,\qquad a<b .
 \label{bq}
\ee
These algebras are called Cayley--Klein (CK)  algebras
\cite{Grom,Gromdos,Sant,Grad}. We will denote them by
$g_{(\k_1, \dots ,\k_N)}$.
Upon rescaling of generators, each $\k_i$ can be reduced separately
to $1$, $0$ or
$-1$, so  the system of CK
algebras  includes $3^N$ Lie algebras.

When all
$\k_i \neq 0$, it is easy to identify (\ref{bp}) with the pseudo-orthogonal
algebra $so(p,q), (p+q=N+1)$ which leaves invariant the bilinear form
\be
\Lambda^{(0)}=\diag(1,\k_{01},\k_{02},\dots,\k_{0N})=
\diag(1,\k_{1},\k_1\k_{2},\dots,\k_1\cdots\k_{N}).
\label{bql}
\ee
Thus for all $\k_i=1$ we recover the $so(N+1)$ algebra. But the
pseudo-orthogonal algebras are far from exhausting the family
of CK algebras. For
instance, if $\k_1= 0$ with all remaining $\k_i\ne 0$ the CK algebra  has a
semidirect sum structure: $g_{(0,\k_2,\dots,\k_N)}=t_N \odot so(p',q')$
$(p'+q'=N)$, where $t_N=\langle J_{0i}\rangle$ $(i=1,\dots,N)$
is an Abelian
subalgebra and the  $so(p',q')=\langle J_{ij}\rangle$ $(i,j=1,\dots,N)$
subalgebra leaves invariant the bilinear form   \be
\Lambda^{(1)}=\diag(1,\k_{12},\k_{13},\dots,\k_{1N})=
\diag(1,\k_{2},\k_2\k_{3},\dots,\k_2\cdots\k_{N}). \label{bqm}
\ee
Therefore, the $N$--dimensional Euclidean  algebra  $iso(N)$ corresponds
to  $g_{(0,1,\dots,1)}$ while the $N$--dimensional Poincar\'e algebra
$iso(N-1,1)$ appears several times: $g_{(0,-1,1,\dots,1)}$,
$g_{(0,-1,-1,1,\dots,1)}$,  $g_{(0,1,-1,\dots,-1,1)}$,
$g_{(0,1,\dots,1,-1)}$,
etc.

Within the set of CK Lie algebras $g_{(\k_1,\dots,\k_N)}$
there exists a family of $N$ In\"on\"u--Wigner (IW) contractions
\be
\phi_{\p_i}(J_{ab})=
\left\{\begin{array}{rl}
\p_i J_{ab}  \     &\mbox{if either $a$ or
$b\in\{0,1,\dots,i-1\}$}\cr
J_{ab}  \     &\mbox{otherwise}
\end{array}\right. ,\quad i=1,\dots,N;
  \label{IWpura}
\ee
that can be applied to any algebra in the CK family. When the
contraction defined by $\phi_{\p_i}$ is performed on
$g_{(\k_1,\dots,\k_i,\dots,\k_N)}$ we get
$g_{(\k_1,\dots,\k_{i-1},0,\k_{i+1},\dots,\k_N)}$  as contracted algebra.
 The set of
contractions just defined   have a commutative character; we can
apply to a given Lie algebra  as many
contractions  of the kind (\ref{IWpura}) as desired and the order
is inmaterial
for the result.

In particular, the contraction
$\phi_{\p_1}$ leads to inhomogeneous algebras and it is called {\it affine
contraction} since the space
$G_{(0,\k_2,\dots,\k_N)}/G_{(\k_2,\dots,\k_N)}$ is a flat affine space.
Another interesting example is the $N$--dimensional Galilean algebra that
corresponds to  $g_{(0,0,1,\dots,1)}$; it can be got from either
the Euclidean or
Poincar\'e algebras by means of the contraction $\phi_{\p_2}$. A explicit
study of CK algebras for  $N=2,3,4$ are given
in refs.\ \cite{BHOS2}, \cite{BHOS3}
and \cite{BH4D}, respectively.

The CK algebras can be viewed from another point of view.
   Let us now consider the algebra $g_{(1,\dots,1)}\equiv so(N+1)$.
   Its non-zero
Lie brackets are
\be
[\JJ_{ab}, \JJ_{ac}] = \JJ_{bc}, \quad
[\JJ_{ab}, \JJ_{bc}] = -\JJ_{ac}, \quad
[\JJ_{ac}, \JJ_{bc}] = \JJ_{ab}, \quad a<b<c.
  \label{bm}
\ee
If we  define the following set of $N$ mappings $\tilde\phi_{\k_i}$
$(i=1,\dots,N)$:  \be
\tilde\phi_{\k_i}(\JJ_{ab})=
\left\{\begin{array}{rl}
\sqrt{\k_i}\JJ_{ab}  \     &\mbox{if either $a$ or
$b\in\{0,1,\dots,i-1\}$}\cr
\JJ_{ab}  \     &\mbox{otherwise}
\end{array}\right. ,\quad \forall \k_i\ne 0;
  \label{Grosimple}
\ee
the
composition of all these mappings gives rise to the formal transformation

\cite{GromTrick}:
\be
J_{ab}:=\Phi_N(\JJ_{ab})=
\tilde\phi_{\k_1}\circ
\cdots
\circ \tilde\phi_{\k_N}(\JJ_{ab})
=\sqrt{\k_{ab}}\, \JJ_{ab},\quad \forall \k_{ab}\ne 0.
 \label{bms}
 \ee
It is easy to check that if the generators $\JJ_{ab}$ close the
$so(N+1)$ algebra
(\ref{bm}), then the transformed generators $J_{ab}$ (all $\k_i$
are different
from zero) close
the CK algebra (\ref{bp}). Thus, strictly speaking,
the transformation (\ref{bms})
relates  $so(N+1)$ with the whole set of pseudo-orthogonal algebras
$so(p,q)$.
Moreover,  if suitably understood as a limiting procedure (whenever some
$\k_i=0$ a limit   $\sqrt{|\k_i|}\to 0$ should be made in the family
$\tilde\phi_{\k_i}$), then (\ref{Grosimple}) and (\ref{bms})
can be formally
applied even when $\k_i$ are allowed to be zero. In this case,
$\Phi_N$ goes
from the $so(N+1)$ algebra to the general CK algebra with no
restrictions as to
the vanishing of the $\k_i$ constants. In this sense, the mapping
(\ref{Grosimple}) is equivalent, when some $\k_i=0$,  to the
combination of a
standard Weyl unitary trick and an IW contraction (\ref{IWpura}),
 the contraction
parameter being  \be \p_i:=\sqrt{|\k_i|},\quad i=1,\dots,N.\label{bqq}
\ee

Therefore, we can profit from the transformation
(\ref{bms}) in a double sense: firstly, {\it any} expression
for the general CK
algebra $g_{(\k_1, \dots \k_N)}$ can be got starting from
the one corresponding
to $so(N+1)$. Secondly, the scheme so obtained automatically
includes a set
of IW contractions implicitly expressed in terms of the $\tilde\phi_{\k_i}$
transformations, as given in (\ref{IWpura}) and (\ref{bqq}).

We apply this device to the contractions of Lie bialgebras according to
the ideas introduced in Sect.\ 2.2. The transformation (\ref{bms})
should be
augmented with the appropriate transformation $\Phi_N(z)$
of the deformation
parameter, in order to obtain a set of LBC's wich can
be applied to {\it any}
bialgebra for the CK algebras. This last step is only required as far as
contractions are involved, because the mere substitution
(\ref{bms}) when applied
to a $so(N+1)$ bialgebra would produce a well defined
coboundary bialgebra for all
pseudo-orthogonal algebras $so(p,q)$ . This procedure gives us
the coboundary and
the fundamental contraction constants that classify the LBC's.

Since we are going to work simultaneously with this
commuting set of classical
contractions
$(\phi_{\p_1},\dots,\phi_{\p_N})$ that will
originate Lie bialgebra ones, it will be useful
to introduce the \begin{em} ``sets of fundamental
bialgebras" \end{em}, that will be denoted by
$(g_{(\k_1,\dots,\k_N)},\delta^{({f_0}_1,\dots,{f_0}_N)})$.
 They will be the result of the  fundamental LBC's
$(\phi_{\p_1},{f_0}_1)$, \dots, $(\phi_{\p_N},{f_0}_N)$
acting on a given original CK bialgebra
    $(g_{(\k_1,\dots,\k_N)},\delta)$ and
give a precise Lie bialgebra for each of the CK algebras.
Hence in this
case, the transformation of the deformation parameter (\ref{bka}) will be
\be
w=\Phi_N(z):=\p_1^{-{f_0}_1}\dots\p_N^{-{f_0}_N}z .
\ee
These sets will underly the quantum algebras denoted by ($U_w
g_{(\k_1,\dots,\k_N)},\Delta)$, which are
deformations of the classical algebra
$g_{(\k_1,\dots,\k_N)}$ in which
$\delta^{({f_0}_1,\dots,{f_0}_N)}$ gives the first order term in $z$ within the
coproduct $\Delta$.  These quantum algebras are straightforwardly
obtained by
applying $\Phi_N$ to $so(N+1)$ in the form (\ref{bld}).

\sect{The so(3) case}

Let us consider the $so(3)$ Lie algebra generated by
$\{\JJ_{01},\JJ_{02},\JJ_{12}\}$. Its standard (uniparametric)
Drinfel'd--Jimbo deformation is given by
the following coproduct $\Delta_{02}$
and commutators $[\ ,\ ]_{02}$:
\bea
&&\co_{02}(\JJ_{02})=1\otimes \JJ_{02} + \JJ_{02}\otimes 1,\nonumber\\
&&\co_{02}(\JJ_{01})=
e^{-\tfrac{z}{2} \JJ_{02}}\otimes \JJ_{01} + \JJ_{01} \otimes
e^{\tfrac{z}{2} \JJ_{02}},\label{ccd}\\
&&\co_{02}(\JJ_{12})=e^{-\tfrac{z}{2} \JJ_{02}}\otimes \JJ_{12} + \JJ_{12}
\otimes e^{\tfrac{z}{2} \JJ_{02}},\nonumber
\eea
\be
[\JJ_{12},\JJ_{01}]_{02}=\frac{\sinh{z \JJ_{02}}}{z}, \qquad
[\JJ_{12},\JJ_{02}]_{02}=-\JJ_{01}, \qquad
[\JJ_{01},\JJ_{02}]_{02}=\JJ_{12}.
\label{cce}
\ee

The label $\{02\}$ in the coproduct and commutation relations reminds the
choice of the primitive generator.

\subsect{$so(3)$ bialgebras and their contractions}

Obviously, we could have written two more
(completely equivalent) structures by choosing
either $\JJ_{01}$ or $\JJ_{12}$
as the primitive elements. Explicitly, these three possibilities are
characterized by the following $r$--matrices:
\be
r_{01}=z(\JJ_{02}\wedge \JJ_{12}),
\quad r_{02}=z(\JJ_{12}\wedge \JJ_{01}),\quad
r_{12}=z(\JJ_{01}\wedge \JJ_{02}).
\label{cd}
\ee
Thus, the associated coboundary bialgebras are given, respectively, by
\bea
&&\delta_{01}(\JJ_{12})=z(\JJ_{12}\wedge \JJ_{01}),\quad
\delta_{01}(\JJ_{01})=0,\quad
\delta_{01}(\JJ_{02})=z(\JJ_{02}\wedge \JJ_{01}); \label{cdc}\\
&&\delta_{02}(\JJ_{12})=z(\JJ_{12}\wedge \JJ_{02}),\quad
\delta_{02}(\JJ_{01})=z(\JJ_{01}\wedge \JJ_{02}),\quad
\delta_{02}(\JJ_{02})=0;  \label{cdd}\\
&&\delta_{12}(\JJ_{12})=0,\quad
\delta_{12}(\JJ_{01})=z(\JJ_{01}\wedge \JJ_{12}),\quad
\delta_{12}(\JJ_{02})=z(\JJ_{02}\wedge \JJ_{12}).  \label{cde}
\eea
 These structures are related among themselves by
means of  appropriate permutations $\pi\in S_3$ on the set of indices
$\{0,1,2\}$:
\be
r_{01}=\pi_{(12)}( r_{02}), \qquad r_{12}=\pi_{(01)}( r_{02}),
\label{cdf}
\ee
and the same for the cocommutators,
where permutations are denoted in cycle
notation, so that $\pi_{(12)}$ is the 2--cycle $(1\, 2)$
on the three indices
$\{0,1,2\}$.

Within $so(3)$, all these Lie bialgebras are equivalent (the intertwining
operator $D$ of Def.\ 2.3 is just defined by the permutation).
However, this
equivalence can be broken when a LBC is performed. If we fix a given set of
Lie algebra contractions (the mappings $\phi_{\p_{i}}$) we find that, in
general, the associated LBC's depend on the coboundary Lie bialgebras
$(so(3),\delta(r_{ab}))$ we started with. In other words:
in $so(3)$ we
have only one (non-degenerate) bialgebra, but for
non-semisimple algebras that
can be obtained as contractions of $so(3)$, in general
there exist more than
one (non-equivalent) Lie bialgebra.

Explicitly, we apply the formal transformation $\Phi_2$ (\ref{bms})
to the
$r$--matrices   and to  the cocommutators of $so(3)$, obtaining in this way

coboundary Lie bialgebras for the $so(3)$ and $so(2,1)$ algebras
included in the
CK algebra $g_{(\k_1,\k_2)}$:  \bea
&& r'_{01}=(\Phi_2^{-1}\otimes\Phi_2^{-1})( r_{01})=
\frac{\Phi_2^{-1}(w)}{\sqrt{\k_1}\k_2}
(J_{02}\wedge J_{12}),\label{chi}\\
&&
r'_{02}=(\Phi_2^{-1}\otimes\Phi_2^{-1})(r_{02})=
\frac{\Phi_2^{-1}(w)}{\sqrt{\k_1\k_2}}
(J_{12}\wedge J_{01}),\label{chj}\\
&& r'_{12}=( \Phi_2^{-1}\otimes\Phi_2^{-1})
(r_{12})= \frac{\Phi_2^{-1}(w)}{\k_1\sqrt{\k_2}}
(J_{01}\wedge J_{02});\label{chk}
\eea
\be
\delta'_{01}(J_{12})=\frac{\Phi_2^{-1}(w)}{\sqrt{\k_1}}(J_{12}\wedge
J_{01}),\quad
\delta'_{01}(J_{01})=0,\quad
\delta'_{01}(J_{02})=\frac{\Phi_2^{-1}(w)}{\sqrt{\k_1}}(J_{02}\wedge
J_{01});\quad \label{cj}
\ee
\be
\delta'_{02}(J_{12})=\frac{\Phi_2^{-1}(w)}{\sqrt{\k_1\k_2}}(J_{12}\wedge
J_{02}),\quad
\delta'_{02}(J_{01})=\frac{\Phi_2^{-1}(w)}{\sqrt{\k_1\k_2}}(J_{01}\wedge
J_{02}),\quad
\delta'_{02}(J_{02})=0;\quad \label{ck}
\ee
\be
\delta'_{12}(J_{12})=0,\quad
\delta'_{12}(J_{01})=\frac{\Phi_2^{-1}(w)}{\sqrt{\k_2}}(J_{01}\wedge
J_{12}),\quad
\delta'_{12}(J_{02})=\frac{\Phi_2^{-1}(w)}{\sqrt{\k_2}}(J_{02}\wedge
J_{12}). \label{cl}
\ee
 The expresions (\ref{chi}--\ref{cl})
contain all the information needed in order to classify the LBC's
we are dealing
with. Let us explain one example. If we choose
$z=\Phi_2^{-1}(w):=\sqrt{\k_1}\, \k_2\, w$, the limits $\k_i\to 0$
of $r'_{01}$
(see (\ref{chi}))  will exist, and the $r$--matrix $r'_{01}=w(J_{02}\wedge
J_{12})$
 will provide a
coboundary bialgebra structure for all the CK algebras
$g_{(\k_1,\k_2)}$. In this
way,   the values for the  coboundary contraction constants
$c_0$  linked to $(g_{(\k_1,\k_2)},\delta'_{01})$ can be obtained. At this
moment, it is important to recall that strictly speaking the contraction
parameter which goes to zero in a contraction limit is indeed

$\p_i=\sqrt{|\k_i|}$, hence in this example the values of $c_0$   are
$1$ for $\phi_{\p_1}$ and $2$ for $\phi_{\p_2}$.
The same
method applied on cocommutators $\delta'_{01}$ gives the fundamental
contraction
constants $f_0$: $1$ for $\phi_{\p_1}$ and $0$ for $\phi_{\p_2}$. The final
result is summarized  in the following theorem:

\noindent {\bf Theorem 3.1.}
\begin{em}
Let $r'_{ab}$ be an $r$--matrix of $g_{(\k_1,\k_2)}$ with
$\k_1,\k_2\ne 0$ and
$\phi_{\p_i}$ the family of automorphisms describing the
classical contraction
  $\k_i\to 0$. The fundamental and the coboundary LBC's
  are defined by the LBC
constants $f_0$ and $c_0$ given in   table I.
\end{em}

\bigskip

{{\bf Table I.} LBC constants for $g_{(\k_1,\k_2)}$.}
\smallskip

\begin{tabular}{|c|cc|cc|}
\hline
Lie  & \multicolumn{2}{c|}{ $\phi_{ {\p_1}} $ }  &
 \multicolumn{2}{c|}{ $\phi_{ {\p_2}}$ } \\
\cline{2-5}
 bialgebra & \  $f_0$ & $c_0$\   & \   $f_0$ & $c_0$ \  \\
\hline
\   $(g_{(\k_1 , \k_2)} , \delta'_{01})$ \     & \  $ 1$  &
 $1  $ \   &\    $ 0$  &  $ 2  $\   \\
\hline
$(g_{(\k_1, \k_2)},  \delta'_{02})$   & \  $ 1$ &  $ 1 $ \
& \   $ 1 $ &  $1 $ \   \\
\hline
$(g_{(\k_1 , \k_2)} , \delta'_{12})$   &\   $ 0 $ &   $2 $ \   & \  $
1$ &  $ 1 $   \  \\
\hline
\end{tabular}

\bigskip

This approach  gives an unified overview of the many contracted
structures in a
very condensed way. Some of these structures are new and others
appear  in the
literature.   For instance, let us study the affine contraction
$\phi_{\p_1}$ which carries $so(3)$ into the Euclidean
algebra. When we choose not to transform the deformation parameter
under this contraction (LBC $(\phi_{ {\p_1}},0)$)
we have $z=w({\k_1})^0=w$,
and $r'_{12}$ does not exist, but the contracted
cocommutator $\delta'_{12}$
does. This case   corresponds
 to the non-coboundary  bialgebra underlying
Vaksman-Korogodski $e(2)_q$ algebra \cite{Kor} (with $\k_2=1$),
which is in turn
dual to Woronowicz's $E(2)_q$ group \cite{Woro}. In this case,   when
defined, LBC's which do not change the deformation parameter ($z=w$)
give {\it only} non-coboundary Lie bialgebras, as
the contractions studied in
\cite{Grom2}. On the contrary, if we allow the
transformation of the parameter
$z$ corresponding to the LBC
$(\phi_{ {\k_1}},2)$, then we find that $r'_{12}$ exists
under $\k_1\to 0$, but $\delta'_{12}$ becomes trivial
in this circumstance (and, hence, the Lie
bialgebra is trivial): as we shall see immediately, the Hopf algebra
contraction so defined leads to a non-deformed structure.

\subsect{Hopf algebra contractions of $U_z so(3)$}

\noindent Table I provides three sets of
fundamental bialgebras:
$ (g_{(\k_1,\k_2)},\delta_{01}^{(1,0)})$,

$(g_{(\k_1,\k_2)},\delta_{02}^{(1,1)})$ and

$(g_{(\k_1,\k_2)},\delta_{12}^{(0,1)})$.
This means that, for each of the nine CK algebras with fixed $\k_1$ and
$\k_2$, we have {\it three} --not necessarily equivalent-- Lie bialgebra
structures. These three sets of fundamental
LBC's define three sets of Hopf algebra
contractions that give rise to three quantum deformations of the CK
algebra $g_{(\k_1,\k_2)}$:

\noindent (1) The fundamental LBC's   defined by $(g_{(\k_1,\k_2)},
\delta_{02}^{(1,1)})$ leads to the choice of the
transformation law for the deformation parameter
($z=\sqrt{\k_1\k_2}w$) taken
originally for the CK scheme in \cite{BHOS2}. In this case,
the quantum algebra
$U_w g_{(\k_1,\k_2)}$ has coproduct $\co_{02}$ (\ref{ccd})
and the following commutation rules
\be
[J_{12},J_{01}]_{02}=\frac{\sinh{w
J_{02}}}{w}, \qquad  [J_{12},J_{02}]_{02}=-\k_2J_{01}, \qquad
[J_{01},J_{02}]_{02}=\k_1 J_{12}.
\label{cle}
\ee
The Hopf algebras $(U_w g_{(0,1)},\Delta_{02})$,
$(U_w g_{(0,-1)},\Delta_{02})$
and $(U_w g_{(0,0)},\Delta_{02})$ correspond, in this order, to quantum
deformations of the Euclidean, Poincar\'e and Galilean
(Heisenberg) algebras and
have been studied in  \cite{CGsu2},  \cite{BCGpho} and
\cite{CGh1}, respectively.
Note also that $(g_{(\k_1,\k_2)},\delta_{02}^{(1,1)})$ is
the only  set of fundamental   bialgebras  which are coboundary as well.

\noindent (2) The Hopf algebra that quantizes $ (g_{(\k_1,\k_2)},
\delta_{01}^{(1,0)})$ is obtained in two steps: firstly, we apply
the permutation $\pi_{(12)}$ on (\ref{ccd}--\ref{cce}).
Afterwards, the LBC's
associated to this case are evaluated obtaining:
\bea
&&\co_{01}(J_{01})=1\otimes J_{01} + J_{01}\otimes 1,\nonumber\\
&&\co_{01}(J_{02})=e^{-\tfrac{w}{2} J_{01}}\otimes J_{02} + J_{02}
\otimes e^{\tfrac{w}{2} J_{01}},\label{clf}\\
&&\co_{01}(J_{12})=e^{-\tfrac{w}{2} J_{01}}\otimes J_{12} + J_{12}
\otimes e^{\tfrac{w}{2} J_{01}},\nonumber \eea
\be
[J_{12},J_{01}]_{01}=J_{02}, \qquad   [J_{12},J_{02}]_{01}=-\k_2
\frac{\sinh{w J_{01}}}{w}, \qquad
[J_{01},J_{02}]_{01}=\k_1 J_{12}.
\label{clg}
\ee
Now, the Hopf algebras $(U_w g_{(\k_1,0)},
\Delta_{01})$ have classical commutation rules and no
$r$--matrix.

\noindent (3) Finally, the same method applied to

$(g_{(\k_1,\k_2)},\delta_{12}^{(0,1)})$ gives
\bea
&&\co_{12}(J_{12})=1\otimes J_{12} + J_{12}\otimes 1,\nonumber\\
&&\co_{12}(J_{01})=e^{-\tfrac{w}{2} J_{12}}\otimes J_{01} + J_{01}
\otimes e^{\tfrac{w}{2} J_{12}},\label{clh}\\
&&\co_{12}(J_{02})=e^{-\tfrac{w}{2} J_{12}}\otimes J_{02} + J_{02}
\otimes e^{\tfrac{w}{2} J_{12}},\nonumber
\eea
\be
[J_{12},J_{01}]_{12}=J_{02}, \qquad   [J_{12},J_{02}]_{12}=-\k_2
J_{01} , \qquad
[J_{01},J_{02}]_{12}=\k_1 \frac{\sinh{w J_{12}}}{w}.
\label{cli}
\ee
The affine contraction $\k_1\to 0$ leads to non-coboundary deformations.
In
particular, if $\k_2=-1$ we obtain a  quantum
(1+1) Poincar\'e algebra analogous to the Vaksman--Korogodsky
deformation of
the Euclidean case $(\k_2=1)$.

On the other hand,  as first order terms within a Hopf
algebra deformation, the bialgebras here described give
information about the commutation rules in
the dual Hopf algebras that define the quantum groups linked
with our algebras. If the cocommutator is written as:
\be
\delta'
(X_\gamma)= f_\gamma^{\alpha \beta} X_\alpha\otimes X_\beta,\label{cm}
\ee
then the dual bracket is (first
order in $w$ and linear in
the generators):
\be
[x^\alpha,x^\beta]=f_\gamma^{\alpha \beta}
x^\gamma ,\label{cn}
\ee
plus, in general, higher order terms both in
$w$ and in the generators. Therefore, these three sets of
fundamental  bialgebras give rise to bialgebras whose
``linearized" dual algebras are characterized by the
brackets given in table II. Note that these ``first
order dual algebras" are solvable ones and, in this
case, no dual bracket depends on the contraction
parameters. It is worth to recall that, at the classical level,
these brackets are the first order terms of the Sklyanin
ones (either generated by a given $r$--matrix or not)
that provide the Poisson-Lie structures whose
deformation-quantization (see \cite{Tak}) gives the quantum groups
mentioned above.

\bigskip

\noindent
{{\bf Table II.} Dual commutators of fundamental bialgebras.}
\smallskip

\noindent
\begin{tabular}{|c|c|c|c|}
\hline  Dual bracket &
$Fun_w(g_{(\k_1,\k_2)},\delta_{01}^{(1,0)})$   &
$Fun_w(g_{(\k_1,\k_2)},\delta_{02}^{(1,1)})$
 & $Fun_w(g_{(\k_1,\k_2)},\delta_{12}^{(0,1)})$\\
\hline
\quad$[j_{12},j_{01}]$\quad & $ w\, j_{12}$ & $0$  & $- w\, j_{01}$ \\
\hline
\quad$[j_{12},j_{02}]$\quad & $0$ & $w\, j_{12}$  & $- w\, j_{02}$ \\
\hline
\quad$[j_{01},j_{02}]$\quad & $-w\, j_{02}$ & $ w\, j_{01}$  & $0$ \\
\hline
\end{tabular}

\bigskip

Finally, the existence of a quantum $R$--matrix for the contracted
Hopf algebra
can be easily explored by using the LBC approach.
If we want $R=1+r+\dots$ to
be a solution of the quantum YBE, then $r$ has to be a
solution for its classical counterpart: $[r,r]=0$.
For instance, it can be
checked that the element
\be
\tilde r_{02}=z\,  (J_{12}\wedge J_{01} + i
(J_{01}\otimes J_{01} + J_{02}\otimes
J_{02}+J_{12}\otimes J_{12}))\label{cp}
\ee
generates the same $so(3)$ coboundary bialgebra as
$r_{02}$ and fulfills the
classical YBE. In fact, the $r$--matrix (\ref{cp})
is the first order term of
the universal $R$--matrix corresponding to $(U_wso(3),\Delta_{02})$. By
applying $\Phi_2$ on it we obtain
\be
\tilde r'_{02}=\frac{\Phi_2^{-1}(z)}{\k_1\k_2}
(\sqrt{\k_1\k_2}J_{12}\wedge
J_{01} +i(\k_2 J_{01}\otimes J_{01} + J_{02}\otimes J_{02}
 +\k_1 J_{12}\otimes
J_{12})) .\label{cq}
\ee

We see that the LBC's for $(so(3),\delta(\tilde r_{02}))$
characterized in
table I do not provide well defined expressions for
the limits $\k_i\to 0$ of
(\ref{cq}). We would have to take $z=\k_1\k_2\, w$
in order to get a contraction
of $\tilde r_{02}$, but in that case both LBC's are
not fundamental and give
the trivial bialgebra as  a contracted structure.
This fact is related to the
problem of obtaining a universal $R$--matrix
verifying the quantum YBE  for the
Euclidean or Poincar\'e algebras by contraction;
the LBC analysis shows that,
for $e(2)$ or $p(1+1)$ this is not possible.

\sect{The so(4) case}

We now start from the $so(4)$ classical $r$--matrix given by
\be
r_{03,12}=z(\JJ_{13}\wedge \JJ_{01} + \JJ_{23}\wedge \JJ_{02}).
\label{dd}
\ee
The coboundary bialgebra $(so(4),\delta(r_{03,12}))$ is, explicitly,
\bea
&& \delta_{03,12}(\JJ_{03})=\delta_{03,12}(\JJ_{12})=0,\nonumber\\
&& \delta_{03,12}(\JJ_{01})=z( \JJ_{01}\wedge \JJ_{03} + \JJ_{23}\wedge
\JJ_{12}),\nonumber\\
&& \delta_{03,12}(\JJ_{02})= z(
\JJ_{02}\wedge \JJ_{03} - \JJ_{13}\wedge
\JJ_{12}),\label{dddc}\\
&& \delta_{03,12}(\JJ_{13})=z(\JJ_{13}\wedge
\JJ_{03} -  \JJ_{02}\wedge \JJ_{12}),\nonumber\\
&& \delta_{03,12}(\JJ_{23})= z(
\JJ_{23}\wedge \JJ_{03} +  \JJ_{01}\wedge \JJ_{12}).
\nonumber
\eea
The label $\{03,12\}$ recalls the pair
of generators which are primitive, i.e.,
their cocommutators are zero. The first
pair of subscripts indicates the ``principal" primitive
generator which appears
in the   term of the cocommutator $\delta(X)$ containing
$X$ and the second one
labels the ``secondary" primitive generator that  generates the quantum
$so(2)$ subalgebra  (now unidimensional) embedded into $so(4)$.

The quantum $so(4)$ algebra linked to this specific bialgebra
is given by the
following coproducts and deformed commutation rules (omitting the labels
$\{03,12\}$) \cite{BHOS3}:
\bea
&&\co(\JJ_{03})=1\otimes \JJ_{03} + \JJ_{03}\otimes
1,\qquad \co(\JJ_{12})=1\otimes \JJ_{12} + \JJ_{12}\otimes
1,\nonumber\\  &&\co(\JJ_{01})= e^{-{z\over
2}\JJ_{03}}\cosh(\tfrac z 2 \JJ_{12}) \otimes \JJ_{01} +
\JJ_{01} \otimes e^{{z\over2}\JJ_{03}} \cosh(\tfrac z 2
\JJ_{12})\nonumber\\ &&\qquad -e^{-{z\over
2}\JJ_{03}}\sinh(\tfrac z 2 \JJ_{12}) \otimes \JJ_{23} +
\JJ_{23}\otimes e^{{z\over2}\JJ_{03}}\sinh(\tfrac z 2
\JJ_{12}),\nonumber\\ &&\co(\JJ_{02})=e^{-{z\over
2}\JJ_{03}}\cosh(\tfrac z 2 \JJ_{12}) \otimes \JJ_{02} +
\JJ_{02} \otimes e^{{z\over2}\JJ_{03}} \cosh(\tfrac z 2
\JJ_{12})\nonumber\\ &&\qquad + e^{-{z\over
2}\JJ_{03}}\sinh(\tfrac z 2 \JJ_{12}) \otimes \JJ_{13} -
\JJ_{13}\otimes e^{{z\over2}\JJ_{03}}\sinh(\tfrac z 2
\JJ_{12}),\label{dddf}\\ &&\co(\JJ_{13})=e^{-{z\over
2}\JJ_{03}}\cosh(\tfrac z 2 \JJ_{12}) \otimes \JJ_{13} +
\JJ_{13} \otimes e^{{z\over2}\JJ_{03}} \cosh(\tfrac z 2
\JJ_{12})\nonumber\\ &&\qquad + e^{-{z\over
2}\JJ_{03}}\sinh(\tfrac z 2 \JJ_{12}) \otimes \JJ_{02} -
\JJ_{02} \otimes e^{{z\over2}\JJ_{03}}\sinh(\tfrac z 2
\JJ_{12}),\nonumber\\ &&\co(\JJ_{23})=e^{-{z\over
2}\JJ_{03}}\cosh(\tfrac z 2 \JJ_{12}) \otimes \JJ_{23} +
\JJ_{23} \otimes e^{{z\over2}\JJ_{03}} \cosh(\tfrac z 2
\JJ_{12})\nonumber\\ &&\qquad -e^{-{z\over
2}\JJ_{03}}\sinh(\tfrac z 2 \JJ_{12}) \otimes \JJ_{01} +
\JJ_{01}\otimes e^{{z\over2}\JJ_{03}}\sinh(\tfrac z 2 \JJ_{12});
\nonumber \eea
\bea
&&\back\back\back [\JJ_{13},\JJ_{23}]=\frac 1z
\sinh(z \JJ_{12})\cosh(z \JJ_{03}),\quad
[\JJ_{23},\JJ_{02}]=\frac 1z
\sinh(\JJ_{03})\cosh(z\JJ_{12}),\nonumber  \\
&&\back\back\back [\JJ_{13},\JJ_{01}]=\frac 1z
\sinh(z\JJ_{03})\cosh(z\JJ_{12}),\quad
[\JJ_{01},\JJ_{02}]=\frac 1z \sinh(z\JJ_{12})\cosh(z\JJ_{03}).
\label{dddg}
\eea

\subsect{$so(4)$ bialgebras and their contractions}

We can build five more
coboundary bialgebras by permutation of the set of indices $\{0,1,2,3\}$.
For
instance, the transition  $r_{03,12}\rightarrow r_{01,23}$ can be got by
applying  on the former
$r$--matrix any of the
permutations mapping 03 into 01 and 12 into 23. There are four such
permutations with cycle decomposition:
$\pi_{(123)},\pi_{(13)},\pi_{(0123)}$ and
$\pi_{(013)}$; all of them provide the same result. In particular we choose the
following representatives for these five structures:
\bea
&& r_{01,23}=\pi_{(13)} (r_{03,12}), \qquad
r_{02,13}=\pi_{(23)} (r_{03,12}), \qquad
r_{12,03}=\pi_{(01)(23)} (r_{03,12}),\nonumber\\
&& r_{13,02}=\pi_{(01)} (r_{03,12}), \qquad
r_{23,01}=\pi_{(02)} (r_{03,12}) .
\label{ddc}
\eea
Similarly, we can deduce their corresponding cocommutators.

The formal
transformation $\Phi_3$
(\ref{bms})  applied to the six $r$--matrices
(\ref{dd},\ref{ddc}) and to their corresponding cocommutators,
gives rise to
coboundary Lie bialgebras for the eight CK algebras $g_{(\k_1,\k_2,\k_3)}$
 with all $\k_i\ne 0$:
$so(4)$, $so(3,1)$ (4 times) and $so(2,2)$ (3 times). Explicit expressions
are given in Appendix A.

Therefore,  it is straightforward to obtain the classification of the
LBC's for each one of the three classical contractions $\phi_{\p_i}$
$(i=1,2,3)$: it suffices to study the properties of
the limit where just one
 $\k_i$ ($\p_i$) goes to zero while keeping the
other two unchanged. Afterwards we have to
find for each case, the fundamental
contraction constant $f_0$ (the minimum value of $n$ which guarantees that
$\delta'=\lim_{\p_i\to 0}\delta$ exists), and  the coboundary
contraction constant $c_0$
(idem with respect to $r'=\lim_{\p_i\to 0}r$). We
state the final result as:

\noindent {\bf Theorem 4.1.}
\begin{em}
Consider the coboundary Lie bialgebra
$(g_{(\k_1,\k_2,\k_3)},\delta'(r'_{ab,cd}))$  $\forall \k_i\ne 0$.  The
fundamental and the coboundary LBC's are defined, in this order,  by the
contraction constants $f_0$ and $c_0$ associated to a given classical
contraction $\phi_{\p_i}$ displayed in   table III.
\end{em}

\newpage

{{\bf Table III.} LBC constants for
$g_{(\k_1,\k_2,\k_3)}$.}
\smallskip

\begin{tabular}{|c|cc|cc|cc|}
\hline
Lie  & \multicolumn{2}{c|}{ $\phi_{ {\p_1}} $ }  &
 \multicolumn{2}{c|}{ $\phi_{ {\p_2}}$ }  &
 \multicolumn{2}{c|}{ $\phi_{ {\p_3}}$ }\\
\cline{2-7}
 bialgebra & \  $f_0$ & $c_0$\   & \   $f_0$ & $c_0$ \ &
 \   $f_0$ & $c_0$ \  \\
\hline
\ $(g,\delta'_{01, 23})$ \  &\
$ 1 $ & $    1 $\  &
\ $ 0 $ & $    2 $\  &\
$ 2 $ & $    2$\  \\
\hline
\ $(g,\delta'_{02,13})$\  &
\ $ 1 $ & $    1 $\  &\
$ 1 $ & $    1 $\  &
\ $ 2 $ & $    2 $\  \\
\hline
\ $(g,\delta'_{03, 12})$\  &\
$ 1 $ & $   1 $\  &\
$ 1 $ & $   1 $\  &\
$ 1 $ & $    1 $\  \\
\hline
\ $(g,\delta'_{12, 03})$\  &\
$ 2 $ & $    2 $\  &\
$ 1 $ & $    1 $\  &\
$ 2 $ & $    2 $\  \\
\hline
\ $(g,\delta'_{13,  02})$\  &\
$ 2 $ & $    2 $\  &\
$ 1 $ & $    1 $\  &\
$ 1 $ & $    1 $\  \\
\hline
\ $(g,\delta'_{23, 01})$\  &\
$ 2 $ & $    2 $\  &\
$ 0 $ & $    2 $ \ &\
$ 1 $ & $    1 $\  \\
\hline
\end{tabular}

\bigskip

Before going to the quantum algebras note that, once again,
all fundamental
LBC's that preserve the deformation parameter ($f_0=0$)
are not of the coboundary
type. In general, non-coboundary bialgebras arise
whenever $f_0<c_0$, this is,
when the LBC $(\phi_{{\p_2}},0)$ is applied on bialgebras that come from
$(so(4),\delta(r_{01,23}))$ and $(so(4),\delta(r_{23,01}))$. It is
easy to check that, if $\k_2\to 0$, no antisymmetric element
of $g\otimes g$
can generate the contracted cocommutators (\ref{dl}) or (\ref{dq}).
This agrees
with Zakrzewski's theorem \cite{Zak} proving that,
for $N>2$, all semidirect
sums $t_N \odot so(p',q')$ with $p'+q'=N$ do not admit non-coboundary
bialgebras (in our case,
the algebras with $\k_2=0$ do not admit such a kind of semidirect sum
decomposition). Among this non-coboundary objects,
we can find two different
(2+1) Galilean  bialgebras: $(g_{(0,0,1)},\delta_{01,23}^{(1,0,2)})$ and
$(g_{(0,0,1)},\delta_{23,01}^{(2,0,1)})$, that we shall relate in the
sequel with two different quantum (2+1) Galilean algebras.

\subsect{Hopf algebra contractions of $U_z so(4)$}

We have obtained six sets of fundamental
bialgebras, and  four of them are fundamental coboundary ones.
Among them, the Hopf structure of

$(g_{(\k_1,\k_2,\k_3)},\delta_{03,12}^{(1,1,1)})$ has been studied in
\cite{BHOS3}; the particular cases
$(U_wg_{(-1,-1,1)},\Delta_{03,12})$ $(so(3,1)_q)$ and

$(U_wg_{(0,1,1)},\Delta_{03,12})$ $(e(3)_q)$ have been given
in \cite{Lukisub}
and \cite{CGe3}, respectively. We shall explicitly discuss two examples
 containing non-coboundary objects, since they provide new
quantizations of some interesting algebras.

\noindent $\bullet$
By applying the permutation $\pi_{(13)}$ to relations
(\ref{dddf}--\ref{dddg}) and the Hopf algebra contractions defined
by the set of fundamental LBC's given in the
first row of table III, we obtain
the Hopf algebra $(U_w g_{(\k_1,\k_2,\k_3)},\co_{01,23})$
(we omit the $\{01,23\}$ label):
\bea
&&\co(J_{01})=1\otimes J_{01} + J_{01}\otimes 1,
\qquad \co(J_{23})=1\otimes J_{23} + J_{23}\otimes
1,\nonumber\\
&&\co(J_{02})= e^{-{w\over
2}\k_3J_{01}}\Ck_{-\k_1\k_3}(\tfrac{w}{2}J_{23}) \otimes J_{02}
+ J_{02} \otimes e^{{w\over2}\k_3J_{01}}
\Ck_{-\k_1\k_3}(\tfrac{w}{2}J_{23})\nonumber\\
&&\qquad +e^{-{w\over
2}\k_3J_{01}}\Sk_{-\k_1\k_3}(\tfrac{w}{2}J_{23}) \otimes \k_1 J_{13} -
\k_1J_{13}\otimes
e^{{w\over2}\k_3J_{01}}\Sk_{-\k_1\k_3}(\tfrac{w}{2}J_{23}),\nonumber\\
&&\co(J_{03})= e^{-{w\over
2}\k_3J_{01}}\Ck_{-\k_1\k_3}(\tfrac{w}{2}J_{23}) \otimes J_{03}
+ J_{03} \otimes e^{{w\over2}\k_3J_{01}}
\Ck_{-\k_1\k_3}(\tfrac{w}{2}J_{23})\nonumber\\
&& - e^{-{w\over
2}\k_3J_{01}}\Sk_{-\k_1\k_3}(\tfrac{w}{2}J_{23}) \otimes \k_1\k_3 J_{12} +
\k_1\k_3J_{12}\otimes
e^{{w\over2}\k_3J_{01}}\Sk_{-\k_1\k_3}(\tfrac{w}{2}J_{23}),\label{co1}\\
&&\co(J_{12})= e^{-{w\over
2}\k_3J_{01}}\Ck_{-\k_1\k_3}(\tfrac{w}{2}J_{23}) \otimes J_{12}
+ J_{12} \otimes e^{{w\over2}\k_3J_{01}}
\Ck_{-\k_1\k_3}(\tfrac{w}{2}J_{23})\nonumber\\
&&\qquad - e^{-{w\over
2}\k_3J_{01}}\Sk_{-\k_1\k_3}(\tfrac{w}{2}J_{23}) \otimes J_{03} +
J_{03}\otimes
e^{{w\over2}\k_3J_{01}}\Sk_{-\k_1\k_3}(\tfrac{w}{2}J_{23}),\nonumber\\
&&\co(J_{13})= e^{-{w\over
2}\k_3J_{01}}\Ck_{-\k_1\k_3}(\tfrac{w}{2}J_{23}) \otimes J_{13}
+ J_{13} \otimes e^{{w\over2}\k_3J_{01}}
\Ck_{-\k_1\k_3}(\tfrac{w}{2}J_{23})\nonumber\\
&&\qquad +e^{-{w\over
2}\k_3J_{01}}\Sk_{-\k_1\k_3}(\tfrac{w}{2}J_{23}) \otimes \k_3 J_{02} -
\k_3J_{02}\otimes
e^{{w\over2}\k_3J_{01}}\Sk_{-\k_1\k_3}(\tfrac{w}{2}J_{23});\nonumber
\eea
\bea
&&[J_{02},J_{03}]= \k_1\k_2\frac 1w \Sk_{-\k_1\k_3 }(wJ_{23})\Ck_{-\k_3^2
}(wJ_{01}),\nonumber\\
&&[J_{02},J_{12}]=\k_2\frac 1w\Sk_{-\k_3^2}(wJ_{01})\Ck_{- \k_1\k_3
}(wJ_{23}),\nonumber\\
&& [J_{12},J_{13}]=\k_2\frac 1w\Sk_{-\k_1\k_3
}(wJ_{23})\Ck_{- \k_3^2 }(wJ_{01}),\label{bra1}\\
&& [J_{03},J_{13}]=\k_2\k_3\frac 1w\Sk_{- \k_3^2}(wJ_{01})\Ck_{-  \k_1\k_3
}(wJ_{23}).\nonumber
\eea
We recall \cite{BHOS2} that
\be
\Ck_{-\k}(x):=\frac{e^{\sqrt{\k} x}+e^{-\sqrt{\k}
x}}2,\qquad \Sk_{-\k}(x):=\frac{e^{\sqrt{\k} x}- e^{-\sqrt{\k}
x}}{2\sqrt{\k}}.
\ee
It is important to point out that in this case the contractions
$\k_2\to 0$ or $\k_3\to 0$ give a deformed coproduct with classical
commutation
rules. We shall insist in this point later.

\noindent $\bullet$ The Hopf algebra $(U_w
g_{(\k_1,\k_2,\k_3)},\co_{23,01})$ is given (omitting the label again) by
\bea
&&\co(J_{23})=1\otimes J_{23} + J_{23}\otimes 1,
\qquad \co(J_{01})=1\otimes J_{01} + J_{01}\otimes
1,\nonumber\\
&&\co(J_{02})= e^{-{w\over
2}\k_1J_{23}}\Ck_{-\k_1\k_3}(\tfrac{w}{2}J_{01}) \otimes J_{02}
+ J_{02} \otimes e^{{w\over2}\k_1J_{23}}
\Ck_{-\k_1\k_3}(\tfrac{w}{2}J_{01})\nonumber\\
&&\qquad + e^{-{w\over
2}\k_1J_{23}}\Sk_{-\k_1\k_3}(\tfrac{w}{2}J_{01}) \otimes \k_1 J_{13} -
\k_1J_{13}\otimes
e^{{w\over2}\k_1J_{23}}\Sk_{-\k_1\k_3}(\tfrac{w}{2}J_{01}),\nonumber\\
&&\co(J_{03})= e^{-{w\over
2}\k_1J_{23}}\Ck_{-\k_1\k_3}(\tfrac{w}{2}J_{01}) \otimes J_{03}
+ J_{03} \otimes e^{{w\over2}\k_1J_{23}}
\Ck_{-\k_1\k_3}(\tfrac{w}{2}J_{01})\nonumber\\
&& - e^{-{w\over
2}\k_1J_{23}}\Sk_{-\k_1\k_3}(\tfrac{w}{2}J_{01}) \otimes \k_1\k_3 J_{12}
+ \k_1\k_3J_{12}\otimes
e^{{w\over2}\k_1J_{23}}\Sk_{-\k_1\k_3}(\tfrac{w}{2}J_{01}),\label{co2}\\
&&\co(J_{12})= e^{-{w\over
2}\k_1J_{23}}\Ck_{-\k_1\k_3}(\tfrac{w}{2}J_{01}) \otimes J_{12}
+ J_{12} \otimes e^{{w\over2}\k_1J_{23}}
\Ck_{-\k_1\k_3}(\tfrac{w}{2}J_{01})\nonumber\\
&&\qquad  - e^{-{w\over
2}\k_1J_{23}}\Sk_{-\k_1\k_3}(\tfrac{w}{2}J_{01}) \otimes J_{03} +
J_{03}\otimes
e^{{w\over2}\k_1J_{23}}\Sk_{-\k_1\k_3}(\tfrac{w}{2}J_{01}),\nonumber\\
&&\co(J_{13})= e^{-{w\over
2}\k_1J_{23}}\Ck_{-\k_1\k_3}(\tfrac{w}{2}J_{01}) \otimes J_{13}
+ J_{13} \otimes e^{{w\over2}\k_1J_{23}}
\Ck_{-\k_1\k_3}(\tfrac{w}{2}J_{01})\nonumber\\
&&\qquad + e^{-{w\over
2}\k_1J_{23}}\Sk_{-\k_1\k_3}(\tfrac{w}{2}J_{01}) \otimes \k_3 J_{02} -
\k_3J_{02}\otimes
e^{{w\over2}\k_1J_{23}}\Sk_{-\k_1\k_3}(\tfrac{w}{2}J_{01});\nonumber
\eea
\bea
&&[J_{02},J_{03}]= \k_1\k_2\frac 1w\Sk_{- \k_1^2}(wJ_{23})\Ck_{- \k_1\k_3
}(wJ_{01}),\nonumber\\
&&[J_{02},J_{12}]=\k_2\frac 1w\Sk_{- \k_1\k_3}(wJ_{01})\Ck_{-  \k_1^2
}(wJ_{23}),\nonumber\\
&& [J_{12},J_{13}]=\k_2\frac 1w\Sk_{- \k_1^2
}(wJ_{23})\Ck_{- \k_1\k_3}(wJ_{01}),\label{bra2}\\
&& [J_{03},J_{13}]=\k_2\k_3\frac 1w\Sk_{- \k_1\k_3}(wJ_{01})\Ck_{-  \k_1^2
}(wJ_{23}).\nonumber
\eea
Similarly to the previous example, now  the
coproduct is invariant under the $\k_2\to 0$ limit, and the contractions

$\k_1\to 0$ or $\k_2\to 0$ provide
classical commutation rules. This case can be easily related to the
previous
  one can   by interchanging the role of the
contractions $\phi_{\p_1}$ and $\phi_{\p_3}$.

\subsect{Quantum (2+1) Poincar\'e and Galilei algebras}

The ``geometrical" orthogonal basis we have been working with can
be  interpreted as a kinematical one. In this way we can
explore the differences
underlying these contractions from a physical point of view. In particular, by
assuming the standard kinematical assignation \cite{BHOS3}:
\be
J_{01}=H,\quad J_{02}=P_1,\quad J_{03}=P_2,\quad
J_{12}=K_1,\quad J_{13}=K_2,\quad J_{23}=J,
\label{kin}
\ee
that expresses in a physical basis the Poincar\'e
algebra $g_{(0,-1,1)}$, we obtain six different deformations
(all of coboundary
type since $f_0=c_0$ in all the cases for the
$\phi_{\p_1}$ contraction) that
can be splitted into two classes:

\noindent (1) {\it Quantum (2+1) Poincar\'e algebras with
deformed commutation rules}: the case $(U_w
g_{(0,-1,1)},\co_{01,23})$ is the three-dimensional analogue
to the $\k$--Poincar\'e algebra \cite{Lukap}. Both $(U_w
g_{(0,-1,1)},\co_{02,13})$ and $(U_w
g_{(0,-1,1)},\co_{03,12})$ were studied in \cite{BHOS3}, and can be
interpreted as relativistic symmetries with a discretized
spatial direction.

\noindent (2) {\it Twisted (2+1) Poincar\'e algebras}:
the contraction $\k_1\to
0$ turns commutators of the
 deformations $(U_w g_{(0,-1,1)},\co_{12,03})$,
 $(U_w g_{(0,-1,1)},\co_{13,02})$
and   $(U_w g_{(0,-1,1)},\co_{23,01})$
into classical ones (see (\ref{bra2})). Moreover, the final coproducts
contain only first  order deformations, and four generators are
now primitive  (see (\ref{co2})). As a consequence, these quantum
algebras can be shown to be twistings of the classical structure where the
twisting operator is just the exponential of the classical
$r$--matrix (compare with \cite{Luktwist}). These three structures
appear for a transformation law $z= \k_1\, w$ of the deformation parameter
under the affine contraction, and to our knowdledge,  have not
been so far considered in the literature.
Among them, the quantum algebra $(U_w
g_{(0,-1,1)},\co_{23,01})$ presents interesting properties. Its coproduct
contains the boosts as the non-primitive generators, in the form
\bea
&& \co_{23,01}(K_1)=1\otimes K_1 + K_1 \otimes 1 -
\frac{w}{2} H\wedge P_2,
\nonumber\\
&& \co_{23,01}(K_2)=1\otimes K_2 + K_2 \otimes 1 +
\frac{w}{2} H\wedge P_1,
\label{twist}
\eea
and therefore is a deformation of the (2+1) Poincar\'e algebra
with classical
commutation rules, deformed boosts and rotational
symmetry preserved. All
these properties were required in \cite{Bacry} as
properties for physically
meaningful space-time deformations of Poincar\'e algebra.

As far as the (2+1) Galilei algebra $g_{(0,0,1)}$ is concerned, some new
features appear. We shall have six
deformations obtained from the Poincar\'e ones
by using the LBC's defined by
$\phi_{\p_2}$. This contraction splits into
three classes the structures so
obtained (the kinematical assignation (\ref{kin}) remains the same):

\noindent (1) {\it Quantum (2+1) Galilei algebras with
deformed commutation rules}: there are two cases, $(U_w
g_{(0,0,1)},\co_{02,13})$ and $(U_w
g_{(0,0,1)},\co_{03,12})$. These coboundary deformations
are connected with
the first class of quantum Poincar\'e algebras and were studied in
\cite{BHOS3}.

\noindent (2) {\it Twisted (2+1) Galilei algebras}:
the quantum algebras $(U_w
g_{(0,0,1)},\co_{12,03})$ and $(U_w
g_{(0,0,1)},\co_{13,02})$ are coboundary deformations with classical
commutation rules. The twisting operator is  given by the $r$--matrix.

\noindent (3) {\it Non-coboundary (2+1) Galilei algebras}:
the deformations
$(U_w g_{(0,0,1)},\co_{01,23})$ and $(U_w
g_{(0,0,1)},\co_{23,01})$  have non-deformed commutation rules.
The first one has a deformed coproduct with exponential terms
(it is just the
(2+1) analogue of the  deformation given in \cite{Gill}).
The latter presents a
simpler deformation (with only linear non-primitive terms)
and reproduces the
deformation of boosts (\ref{twist}) in the non-relativistic case.

In this way, we can see how the different
permutations of $so(4)$ bialgebras give
rise to non-equivalent deformations as long as LBC's are applied in a
systematic way.  Finally, let us note that if we construct (as in
table II) the linear part
of the dual commutation rules among the generators
of the dual basis
with respect to a fixed set of fundamental bialgebras, we would find
that here some coefficients $\k_i$ (further to the
contracted deformation
parameter $w$) would appear as structure
constants of the (solvable) dual Lie
algebras; therefore,
  cocommutators are not invariant under fundamental contractions.
However, it is easy to prove that the fundamental
character of the LBC's will
ensure that the Abelian algebra does not appear within
this dual structure.

\sect{The so(5) case}

We summarize the results obtained when this method is applied on
the $so(5)$ coboun-dary bialgebra generated by
\be
r_{04,13}=z(\JJ_{14}\wedge \JJ_{01} + \JJ_{24}\wedge \JJ_{02}
+ \JJ_{34}\wedge \JJ_{03} + \JJ_{23}\wedge \JJ_{12}).
\label{ec}
\ee

A  straightforward computation gives rise to the
following explicit form of the cocommutator $\delta_{04,13}$:
\bea
&& \delta_{04,13}(\JJ_{04})=0,\qquad
\delta_{04,13}(\JJ_{12})=z(\JJ_{12}\wedge
\JJ_{13}),\nonumber\\
&& \delta_{04,13}(\JJ_{13})=0,\qquad
\delta_{04,13}(\JJ_{23})=z(\JJ_{23}\wedge \JJ_{13}),\nonumber\\
&& \delta_{04,13}(\JJ_{02})=z(
\JJ_{02}\wedge \JJ_{04}+
\JJ_{12}\wedge \JJ_{14}+
\JJ_{34}\wedge \JJ_{23}+
\JJ_{23}\wedge \JJ_{01}+
\JJ_{12}\wedge \JJ_{03}), \nonumber\\
&& \delta_{04,13}(\JJ_{24})=z(
\JJ_{24}\wedge \JJ_{04}+
\JJ_{23}\wedge \JJ_{03}+
\JJ_{01}\wedge \JJ_{12}+
\JJ_{12}\wedge \JJ_{34}+
\JJ_{23}\wedge \JJ_{14}), \nonumber\\
&& \delta_{04,13}(\JJ_{01})=z(
\JJ_{01}\wedge \JJ_{04}+
\JJ_{24}\wedge \JJ_{12}+
\JJ_{34}\wedge \JJ_{13}+
\JJ_{02}\wedge \JJ_{23}), \label{ee}\\
&& \delta_{04,13}(\JJ_{34})=z(
\JJ_{34}\wedge \JJ_{04}+
\JJ_{02}\wedge \JJ_{23}+
\JJ_{01}\wedge \JJ_{13}+
\JJ_{24}\wedge \JJ_{12}), \nonumber\\
&& \delta_{04,13}(\JJ_{03})=z(
\JJ_{03}\wedge \JJ_{04}+
\JJ_{13}\wedge \JJ_{14}+
\JJ_{23}\wedge \JJ_{24}+
\JJ_{02}\wedge \JJ_{12}), \nonumber\\
&& \delta_{04,13}(\JJ_{14})=z(
\JJ_{14}\wedge \JJ_{04}+
\JJ_{13}\wedge \JJ_{03}+
\JJ_{12}\wedge \JJ_{02}+
\JJ_{24}\wedge \JJ_{23}). \nonumber
\eea
Note that, in this case,  the term $\JJ_{23}\wedge \JJ_{12}$ in
(\ref{ec}) corresponds to the Lie bialgebra $(so(3),\delta(r_{13}))$
embedded into
$(so(5),\delta(r_{04,13}))$. This Lie subalgebra is generated by $\langle
\JJ_{12},\JJ_{13},\JJ_{23}\rangle$ with  $\JJ_{13}$
playing the role of a
secondary primitive generator.

The 120 elements of the
permutation group $S_5$ on the five indices
$\{0,1,2,3,4\}$ can be casted into 30
different classes attending to their action
on the ordered set of two primitive
generators; each of these classes
includes  four permutations giving rise to
the tetrads $(ab,cd),(ba,cd),(ab,dc)$
and $(ba,dc)$ starting from $(ab,cd)$. The four permutations in each class
would  lead to the same $r$--matrix starting from $r_{04,13}$. Thus, the
mapping $\Phi_4$ (\ref{bms})  leads to 30 Lie bialgebra structures for the
pseudo-orthogonal algebras included in the CK algebra
$g_{(\k_1,\k_2,\k_3,\k_4)}$. The complete classification of
its LBC's is given by the following theorem.

\noindent {\bf Theorem 5.1.}
\begin{em}
LBC's  of the
bialgebra $(g_{(\k_1,\k_2,\k_3,\k_4)},\delta'(r'_{ab,cd}))$ are
classified by the contraction constants given in table IV.
\end{em}

The explicit form of the $so(5)$ quantization corresponding to the
Lie bialgebra (\ref{ee}) is rather complicated, and   has been
studied in \cite{BGHKar}.  However, once the affine
contraction $\k_1\to 0$ is
carried out, the structure is much simpler as it
can be seen, for instance, in the
Hopf algebra that quantizes the coboundary Lie
bialgebra $(g_{(0,\k_2,\k_3,\k_4)},
\delta_{04,13}^{(1,1 ,1,1 )})$ which has been
developed in \cite{BH4D}. Another
interesting examples are the
$\k$--Poincar\'e algebra \cite{Lukap} which appears as a quantization of
$(g_{(0,-1,1,1)}, \delta_{01,34}^{(1,2,2,
2 )})$, and the Giller's (3+1) Galilei
deformation \cite{Gill}   obtained by applying on the former the
$\phi_{\p_2}$ contraction; note that the latter it is
not a coboundary (see third
row of table IV).

These kinematical realizations of the CK algebras are obtained provided a
precise physical meaning for the generators $J_{ab}$
has been considered (see
\cite{BH4D}). This fact gives sense to the exhaustive
classification of LBC's
summarized in the Theorem 5.1. Although table
IV defines 30 sets of quantum CK
algebras  $(g_{(\k_1,\k_2,\k_3,\k_4)}, \Delta_{ab,cd})$, their explicit
expressions are quite cumbersome to obtain. However, first order
deformations are available and provide essential
information to characterize the
deformations. If considered as interesting after
this primary analysis, the whole quantum structure can be straightforwardly
derived by permutation and contraction from the $so(5)$ structure, already
known.

A glimpse on Table IV reveals that the  ratio of fundamental
LBC's   leading to divergencies in the $r$--matrix ($f_0<c_0$) has
diminished strongly with respect to the lower dimensional cases
(no $\phi_{\p_1}$ and $\phi_{\p_4}$ LBC's give such a problem). In fact,
fundamental coboundary contractions ($f_0=c_0$) are the most
frequent objects we obtain, and the contraction constant $f_0=2$
plays now a prominent role.  No twisted (3+1) Poincar\'e algebra preserving
rotational symmetry (similar to (\ref{twist})) seems to be available now.

\newpage

{{\bf Table IV.} LBC constants for
$g_{(\k_1,\k_2,\k_3,\k_4)}$.}
\bigskip

\begin{tabular}{|c|cc|cc|cc|cc|}
\hline
Lie  & \multicolumn{2}{c|}{ $\phi_{ {\p_1}} $ }  &
 \multicolumn{2}{c|}{ $\phi_{ {\p_2}}$ }  &
 \multicolumn{2}{c|}{ $\phi_{ {\p_3}}$ }  &
 \multicolumn{2}{c|}{ $\phi_{ {\p_4}}$ }\\
\cline{2-9}
\ \ bialgebra\ \ \  & \  $f_0$ & $c_0$\   & \   $f_0$ & $c_0$ \ & \
$f_0$ & $c_0$ \
& \   $f_0$ & $c_0$ \  \\
\hline
$(g,\delta'_{01,23})$\quad &
 $ 1$  & $1 $  &
$ 0$ & $2 $  &
$ 2$  & $2 $  &
$2$  & $2 $
\\
\hline
$(g,\delta'_{01,24})$\quad &
$ 1$  & $1 $  &
$ 0$ & $ 2 $  &
$ 2$  & $2 $  &
$ 2$  & $2 $
\\
\hline
$(g,\delta'_{01,34})$\quad &
$ 1$  & $1 $  &
$ 0$ & $ 2 $  &
$ 2$  & $2 $  &
$ 2$  & $2 $
\\
\hline\hline
$(g,\delta'_{02,13})$\quad &
$ 1$  & $1 $  &
$ 1$  & $1 $  &
$ 2$  & $2 $  &
$ 2$  & $2 $
\\
\hline
$(g,\delta'_{02,14})$ &
$ 1$  & $1 $  &
$ 1$  & $1 $  &
$ 2$  & $2 $  &
$ 2$  & $2 $
\\
\hline
$(g,\delta'_{02,34})$ &
$ 1$  & $1 $  &
$ 2$  & $2 $  &
$ 2$  & $2 $  &
$ 2$  & $2 $
\\
\hline
\hline
$(g,\delta'_{12,03})$ &
$ 2$  & $2 $  &
$ 1$  & $1 $  &
$ 2$  & $2$  &
$ 2$  & $2 $
\\
\hline
$(g,\delta'_{12,04})$ &
$ 2$  & $2 $   &
$ 1$  & $1 $  &
$ 2$  & $2 $   &
$ 2$  & $2 $
\\
\hline
$(g,\delta'_{12,34})$ &
$ 2$  & $2 $  &
$ 2$  & $2 $ &
$ 2$  & $2 $   &
$ 2$  & $2 $
\\
\hline\hline
$(g,\delta'_{03,12})$ &
$ 1$  & $1 $  &
$ 1$  & $1 $  &
$ 2$  & $2 $  &
$ 2$  & $2 $
\\
\hline
$(g,\delta'_{03,14})$ &
$ 1$  & $1 $  &
$ 1$  & $1 $  &
$ 1$  & $1 $  &
$ 2$  & $2 $
\\
\hline
$(g,\delta'_{03,24})$ &
$ 1$  & $1 $ &
$ 2$  & $2 $  &
$ 1$  & $1 $  &
$ 2$  & $2 $
\\
\hline
\hline
$(g,\delta'_{13,02})$ &
$ 2$  & $2 $  &
$ 1$  & $1 $  &
$ 2$  & $2 $  &
$ 2$  & $2 $
\\
\hline
$(g,\delta'_{13,04})$ &
$ 2$  & $2 $  &
$ 1$  & $1 $  &
$ 1$  & $1 $  &
$ 2$  & $2 $
\\
\hline
$(g,\delta'_{13,24})$ &
$ 2$  & $2 $  &
$  2$  & $2  $  &
$ 1$  & $1 $  &
$ 2$  & $2 $
\\
\hline
\hline
$(g,\delta'_{23,01})$ &
$ 2$  & $2 $  &
$ 2$  & $2 $  &
$2 $ & $2$ &
$ 2$  & $2 $
\\
\hline
$(g,\delta'_{23,04})$ &
$ 2$  & $2 $  &
$ 2$  & $2 $  &
$ 1$  & $1 $  &
$ 2$  & $2 $
\\
\hline
$(g,\delta'_{23,14})$ &
$ 2$  & $2  $  &
$ 2$  & $2 $  &
$ 1$  & $1 $  &
$ 2$  & $2 $
\\
\hline
\hline
$(g,\delta'_{04,12})$ &
$ 1$  & $1 $  &
$ 1$  & $1 $  &
$ 2$  & $2 $  &
$ 1$  & $1 $
\\
\hline
$(g,\delta'_{04,13})$ &
$ 1$  & $1 $  &
$ 1$  & $1 $  &
$ 1$  & $1 $  &
$ 1$  & $1 $
\\
\hline
$(g,\delta'_{04,23})$ &
$ 1$  & $1 $  &
$ 2$  & $2 $  &
$ 1$  & $1 $  &
$ 1$  & $1 $
\\
\hline
\hline
$(g,\delta'_{14,02})$ &
$ 2$  & $2 $  &
$ 1$  & $1 $  &
$ 2$  & $2 $  &
$ 1$  & $1 $
\\
\hline
$(g,\delta'_{14,03})$ &
$ 2$  & $2 $  &
$ 1$  & $1 $  &
$ 1$  & $1 $  &
$ 1$  & $1 $
\\
\hline
$(g,\delta'_{14,23})$ &
$ 2$  & $2 $  &
$ 2$  & $2 $  &
$ 1$  & $1 $  &
$ 1$  & $1 $
\\
\hline
\hline
$(g,\delta'_{24,01})$ &
$ 2$  & $2 $ &
$ 2$  & $2 $  &
$ 2$  & $2 $  &
$ 1$  & $1 $
\\
\hline
$(g,\delta'_{24,03})$ &
$2$  & $2 $  &
$ 2$  & $2 $  &
$ 1$  & $1 $  &
$ 1$  & $1 $
\\
\hline
$(g,\delta'_{24,13})$ &
$ 2$  & $2 $  &
$2 $ & $2$  &
$ 1$  & $1 $  &
$ 1$  & $1 $
\\
\hline
\hline
$(g,\delta'_{34,01})$ &
$2$  & $2 $  &
$2$  & $2 $  &
$ 0$ & $ 2 $  &
$ 1$  & $1 $
\\
\hline
$(g,\delta'_{34,02})$ &
$ 2$  & $2 $  &
$ 2$  & $2 $  &
$ 0$&$2 $  &
$ 1$  & $1 $
\\
\hline
$(g,\delta'_{34,12})$ &
$ 2$  & $2 $  &
$ 2$  & $2 $ &
$0$ & $  2 $  &
$ 1$  & $1 $
\\
\hline
\end{tabular}

\sect{Concluding remarks}

We have presented a general framework to obtain and classify Lie bialgebra
structures by contraction. Due to its physical interest,
we have developed in a
explicit way the case of quasi-orthogonal algebras endowed
with cocommutators
that support the uniparametric Drinfel'd--Jimbo deformation
for these algebras.
However, this general theory can be straightforwardly
applied to any other Lie
bialgebra.

In order to summarize the information  we have obtained in the
previous sections, it is interesting to emphasize the link
between some quantum
algebras  found in the literature and the underlying Lie bialgebras
  we have obtained by contraction.  We
display   in the following table explicit references   together with
the  notation for the fundamental bialgebra $(g_{(\k_1,\dots\k_N)},
\delta_{ab,cd}^{({f_0}_1,\dots,{f_0}_N)})$ we have used in
this paper (we omit
the constants ${f_0}_i$); for the corresponding quantum
algebras we give a single
reference, no exhaustiveness is attempted here.

\bigskip

\begin{tabular}{|c|c|c|}
\hline
$ \ $Fundamental bialgebra $ \ $&
$ \ $Quantum algebra$\  $ &
$ \ $References$ \ $ \\
\hline
$(g_{(0,1)},\delta_{12})$ & $e(2)_q$ &
\cite{Kor}  \\
\hline
$(g_{(0,1)},\delta_{02})$ & $e(2)_q$ &  \cite{CGsu2} \\
\hline
$(g_{(0,-1)},\delta_{02})$ & $p(1+1)_q$ &  \cite{BCGpho} \\
\hline
$(g_{(0,0)},\delta_{02})$ & $g(1+1)_q$ &  \cite{CGh1} \\
\hline
\hline
$(g_{(0,1,1)},\delta_{03,12})$ & $e(3)_q$ & \cite{CGe3}  \\
\hline
$(g_{(0,-1,1)},\delta_{01,23})$ & $p(2+1)_q$ & \cite{Gill}  \\
\hline
$(g_{(0,-1,1)},\delta_{03,12})$ & $p(2+1)_q$ & \cite{BHOS3}  \\
\hline
$(g_{(0,0,1)},\delta_{01,23})$ & $g(2+1)_q$ & \cite{Gill}  \\
\hline
$(g_{(0,0,1)},\delta_{03,12})$ & $g(2+1)_q$ & \cite{BHOS3}  \\
\hline\hline
$(g_{(0,1,1,1)},\delta_{04,13})$ & $e(4)_q$ & \cite{BH4D}  \\
\hline
$(g_{(0,-1,1,1)},\delta_{01,23})$ & $p(3+1)_q$ & \cite{Lukap}  \\
\hline
$(g_{(0,-1,1,1)},\delta_{04,13})$ & $p(3+1)_q$ & \cite{BH4D}  \\
\hline
$(g_{(0,0,1,1)},\delta_{01,23})$ & $g(3+1)_q$ & \cite{Gill} \\
\hline
$(g_{(0,0,1,1)},\delta_{04,13})$ & $g(3+1)_q$ & \cite{BH4D}  \\
\hline
\end{tabular}

\bigskip

One of the advantages of this systematic approach is the way in
which non-coboundary structures appear (they are natural consequences of
exhausting all contraction possibilities). In general,
the obtention of Lie
bialgebras corresponding to non-semisimple groups is far from being simple,
and the application of LBC's seems to provide a relevant subset among them. In
this classification context, twisted structures appear quite frequently
under contraction.

Another interesting field of applications for this method
is the contraction of
classical Poisson--Lie structures. A first example in this
direction has
been given in \cite{pois}. The multiplicity of Lie bialgebra structures
found in this paper can be translated into Poisson--Lie terms.
The study of LBC's
could serve as a useful tool in order to classify  this kind of Hamiltonian
structures. Work on this line is currently in progress.

\bigskip
\medskip

 \noindent
{\large{{\bf Acknowledgements}}}

\bigskip

N.A.G would like to thank the DGICYT of the Ministerio de Educaci\'on
y  Ciencia  de  Espa\~na for supporting  his sabbatical stay
(SAB93--0134) and  to Valladolid  University for hospitality.
This  work
has been partially supported by the DGICYT from  the  Ministerio  de
Educaci\'on  y  Ciencia  de Espa\~na (projects PB91--0196
and PB92-0255)  and
from the Russian Foundation of Fundamental Research
(project 93-011-16180).

\bigskip
\medskip

\noindent
{\large{\bf{Appendix A: Coboundary Lie bialgebras for
$g_{(\k_1,\k_2,\k_3)}$}}}

\appendix

\setcounter{equation}{0}

\renewcommand{\theequation}{A.\arabic{equation}}

\bigskip

The six coboundary Lie bialgebras for the semi-simple CK algebras
$g_{(\k_1,\k_2,\k_3)}$ with all $\k_i\ne 0$ are obtained
by applying the transformation $\Phi_3$ to the six $r$--matrices
(\ref{dd},\ref{ddc}) and to their corresponding cocommutators.
The analysis of the transformation of the deformation parameter $z=
\Phi_3^{-1}(w)$ provides the fundamental
and coboundary contraction constants displayed in table III.

For instance, consider  the  bialgebra generated by the $r$--matrix:
\be
r'_{01,23}=\frac{\Phi_3^{-1}(w)}{\sqrt{\k_1}\k_2\k_3}( J_{03}\wedge
 J_{13} + \k_3  J_{02}\wedge  J_{12}).\nonumber
 \ee
The cocommutators are:
\bea
&&
\delta'_{01,23}( J_{01})=\delta'_{01,23}( J_{23})=0,\nonumber\\
&&
\delta'_{01,23}( J_{02})=\frac{\Phi_3^{-1}(w)}{\sqrt{\k_1}\k_3}
(\k_3  J_{02}\wedge J_{01} -\k_1  J_{13}\wedge
 J_{23}),\nonumber\\
&&\delta'_{01,23} ( J_{03})=
\frac{\Phi_3^{-1}(w)}{\sqrt{\k_1}\k_3}(\k_3  J_{03}\wedge  J_{01} +
\k_1\k_3  J_{12}\wedge  J_{23}),\label{dl}\\
&&\delta'_{01,23}( J_{12})
=\frac{\Phi_3^{-1}(w)}{\sqrt{\k_1}\k_3}(\k_3  J_{12}\wedge  J_{01} +
 J_{03}\wedge  J_{23}),\nonumber\\
&&\delta'_{01,23} ( J_{13}) =
\frac{\Phi_3^{-1}(w)}{\sqrt{\k_1}\k_3}(\k_3  J_{13}\wedge  J_{01} -
\k_3  J_{02}\wedge  J_{23}).
\nonumber
\eea
We find that the transformations of the deformation
parameter are $z= \sqrt{\k_1}\k_2\k_3\, w$ for the $r$--matrix and
$z=\sqrt{\k_1}\k_3\, w$ for the cocommutators. Hence, attending to the
contraction factors $\p_i$, the coboundary contraction constants are
$c_0=\{1,2,2\}$ while the fundamental contraction constants are
$f_0=\{1,0,2\}$ (we follow the ordering
$\{\phi_{\p_1},\phi_{\p_2},\phi_{\p_3}\}$).

The remaining structures are:

\noindent $\bullet$ Bialgebra generated by
$r'_{02,13}=\frac{\Phi_3^{-1}(w)}{\sqrt{\k_1\k_2}\k_3}(\k_3  J_{12}\wedge
 J_{01} +  J_{03}\wedge  J_{23})$:
\bea
&&
\delta'_{02,13}( J_{02})=\delta'_{02,13}( J_{13})=0,
\nonumber\\
&&
\delta'_{02,13}( J_{01})=\frac{\Phi_3^{-1}(w)}{\sqrt{\k_1\k_2}\k_3}
(\k_3  J_{01}\wedge J_{02} - \k_1  J_{23}\wedge
 J_{13}),\nonumber\\
&&\delta'_{02,13} ( J_{03})=
\frac{\Phi_3^{-1}(w)}{\sqrt{\k_1\k_2}\k_3}( \k_3  J_{03}\wedge  J_{02}
- \k_1\k_3  J_{12}\wedge  J_{13}),\label{dm}\\
&&\delta'_{02,13}( J_{12})
=\frac{\Phi_3^{-1}(w)}{\sqrt{\k_1\k_2}\k_3}(\k_3  J_{12}\wedge  J_{02}
-   J_{03}\wedge  J_{13}),\nonumber\\
&&\delta'_{02,13} ( J_{23}) =
\frac{\Phi_3^{-1}(w)}{\sqrt{\k_1\k_2}\k_3}(\k_3  J_{23}\wedge  J_{02}
- \k_3   J_{01}\wedge  J_{13}).
\nonumber
\eea

\noindent $\bullet$ Bialgebra generated by
$r'_{03,12}=
\frac{\Phi_3^{-1}(w)}
{\sqrt{\k_1\k_2\k_3}}( J_{13}\wedge  J_{01} +  J_{23}\wedge
 J_{02})$:
\bea
&&
\delta'_{03,12}( J_{03})=\delta'_{03,12}( J_{12})=0,\nonumber\\ &&
\delta'_{03,12}( J_{01})=\frac{\Phi_3^{-1}(w)}
{\sqrt{\k_1\k_2\k_3}}( J_{01}\wedge J_{03} +
\k_1 J_{23}\wedge  J_{12}),\nonumber\\
&&\delta'_{03,12} ( J_{02})=
\frac{\Phi_3^{-1}(w)}{\sqrt{\k_1\k_2\k_3}} (
 J_{02}\wedge  J_{03} - \k_1 J_{13}\wedge
 J_{12}),\label{dn}\\
&&\delta'_{03,12}( J_{13})
=\frac{\Phi_3^{-1}(w)}{\sqrt{\k_1\k_2\k_3}}
( J_{13}\wedge  J_{03} - \k_3  J_{02}\wedge
 J_{12}),\nonumber\\
&&\delta'_{03,12} ( J_{23}) =
\frac{\Phi_3^{-1}(w)} {\sqrt{\k_1\k_2\k_3}} (
 J_{23}\wedge  J_{03} + \k_3  J_{01}\wedge  J_{12}).
\nonumber
\eea

\noindent $\bullet$ Bialgebra generated by

$r'_{12,03}=\frac{\Phi_3^{-1}(w)}{\k_1\sqrt{\k_2}\k_3}(\k_3 J_{01}\wedge
 J_{02} + \k_1  J_{13}\wedge  J_{23})$:
\bea
&&
\delta'_{12,03}( J_{03})=\delta'_{12,03}( J_{12})=0,\nonumber\\
&&
\delta'_{12,03}( J_{01})=\frac{\Phi_3^{-1}(w)}{\k_1\sqrt{\k_2}\k_3}
(\k_1\k_3 J_{01}\wedge J_{12} + \k_1  J_{23}\wedge
 J_{03}),\nonumber\\
&&\delta'_{12,03} ( J_{02})=
\frac{\Phi_3^{-1}(w)}{\k_1\sqrt{\k_2}\k_3}( \k_1\k_3 J_{02} \wedge
 J_{12} - \k_1  J_{13}\wedge  J_{03}),\label{do}\\
&&\delta'_{12,03}( J_{13})
=\frac{\Phi_3^{-1}(w)}{\k_1\sqrt{\k_2}\k_3}(\k_1\k_3  J_{13}\wedge
 J_{12} -  \k_3  J_{02}\wedge  J_{03}),\nonumber\\
&&\delta'_{12,03} ( J_{23}) =
\frac{\Phi_3^{-1}(w)}{\k_1\sqrt{\k_2}\k_3} ( \k_1\k_3  J_{23}\wedge
 J_{12} +  \k_3  J_{01}\wedge  J_{03}).
\nonumber
\eea

\noindent $\bullet$ Bialgebra generated by
$r'_{13,02}=\frac{\Phi_3^{-1}(w)}{\k_1\sqrt{\k_2\k_3}}( J_{01}\wedge
 J_{03} + \k_1  J_{23}\wedge  J_{12})$:
\bea
&&
\delta'_{13,02}( J_{13})=\delta'_{13,02}( J_{02})=0,
\nonumber\\
&&
\delta'_{13,02}( J_{01})=\frac{\Phi_3^{-1}(w)}{\k_1\sqrt{\k_2\k_3}}
(\k_1  J_{01}\wedge J_{13} - \k_1  J_{23}\wedge
 J_{02}),\nonumber\\
&&\delta'_{13,02} ( J_{03})=
\frac{\Phi_3^{-1}(w)}{\k_1\sqrt{\k_2\k_3}}( \k_1  J_{03}\wedge  J_{13}
-\k_1 \k_3  J_{12}\wedge  J_{02}),\label{dp}\\
&&\delta'_{13,02}( J_{12})
=\frac{ \Phi_3^{-1}(w)}{\k_1\sqrt{\k_2\k_3}}(\k_1  J_{12}\wedge  J_{13}
-  J_{03}\wedge  J_{02}),\nonumber\\
&&\delta'_{13,02} ( J_{23}) =
\frac{\Phi_3^{-1}(w)}{\k_1\sqrt{\k_2\k_3}} (\k_1  J_{23}\wedge  J_{13}
- \k_3   J_{01}\wedge  J_{02}).
\nonumber
\eea

\noindent $\bullet$ Bialgebra generated by
$r'_{23,01}=\frac{\Phi_3^{-1}(w)}{\k_1\k_2\sqrt{\k_3}}(\k_1  J_{12}\wedge
 J_{13} +  J_{02}\wedge  J_{03})$:
\bea
&&
\delta'_{23,01}( J_{23})=\delta'_{23,01}( J_{01})=0,\nonumber\\
&&
\delta'_{23,01}( J_{02})=\frac{\Phi_3^{-1}(w)}{\k_1\sqrt{\k_3}}
(\k_1  J_{02}\wedge J_{23} -\k_1  J_{13}\wedge
 J_{01}),\nonumber\\
&&\delta'_{23,01} ( J_{03})=
\frac{\Phi_3^{-1}(w)}{\k_1\sqrt{\k_3}}( \k_1  J_{03}\wedge  J_{23} +
\k_1\k_3  J_{12}\wedge  J_{01}),\label{dq}\\
&&\delta'_{23,01}( J_{12})
=\frac{ \Phi_3^{-1}(w)}{\k_1\sqrt{\k_3}}(\k_1  J_{12}\wedge  J_{23} +
 J_{03}\wedge  J_{01}),\cr
&&\delta'_{23,01} ( J_{13}) =
\frac{\Phi_3^{-1}(w)}{\k_1\sqrt{\k_3}}(\k_1  J_{13}\wedge  J_{23} -
\k_3  J_{02}\wedge  J_{01}).
\nonumber
\eea

\end{document}